# STRUCTURAL PROTEIN-BASED FLEXIBLE WHISPERING GALLERY MODE RESONATORS

Huzeyfe Yılmaz[1,*], Abdon Pena-Francesch[2,*], Robert Shreiner[2], Huihun Jung[2], Şahin Kaya Özdemir[1], Lan Yang[1], Melik C. Demirel[2,3]

[1.] Electrical and Systems Engineering Department, Washington University, St. Louis, MO 63130

[2.] Department of Engineering Science and Mechanics, Pennsylvania State University, University Park, PA, 16802

[3.] Materials Research Institute and Huck Institutes of Life Sciences, Pennsylvania State University, University Park, PA, 16802

[*]These authors contributed equally

Corresponding Authors E-mails: LY yang@seas.wustl.edu, ŞKÖ ozdemir@wustl.edu, MCD mdemirel@engr.psu.edu

**Abstract**

Nature provides a set of solutions for photonic structures that are finely tuned, organically diverse and optically efficient. Exquisite knowledge of structure-property relationships in proteins aids in the design of materials with desired properties for building devices with novel functionalities, which are difficult to achieve or previously unattainable. Recent bio-inspired photonic platforms made from proteinaceous materials lay the groundwork for many functional device applications, such as electroluminescence in peptide nucleic acids[1], multiphoton absorption in amyloid fibers[2] and silk waveguides and inverse opals[3]. Here we report whispering-gallery-mode (WGM) microresonators fabricated entirely from semi-crystalline structural proteins (i.e., squid ring teeth, SRT, from *Loligo vulgaris* and its recombinant) with unconventional thermo-optic response. We demonstrated waveguides, add-drop filters and flexible resonators as first examples of energy-efficient, highly flexible, biocompatible and biodegradable protein-based photonic devices. Optical switching efficiency in these devices is over thousand times greater than the values reported for Silica WGM resonators. This work opens the way for designing energy efficient functional photonic devices using structure-property relationships of proteins.

The ability to manipulate and control light flow and light–matter interactions using whispering-gallery-mode (WGM) resonators has created significant interest in various fields of science, including but not limited to biosensing[4] and detection[5], cavity-QED[6], optomechanics[7], and parity-time symmetric photonics[8]. WGM resonators are currently manufactured using microelectronics technologies with conventional materials such as silica[9], silicon[10], and silicon nitride[11]. However, recent developments in optical technologies have revealed the strong need for developing soft, biocompatible, and biodegradable photonic devices and photonic structures with novel functionalities that cannot be attained with current optical materials[12]. Towards this aim, all-polymer[13] and polymer-coated silica WGM resonators[14], as well as silica WGM resonators encapsulated in low-index polymers[15] have been fabricated using conventional and commercially available polymers, such as polydimethylsiloxane (PDMS) and polystyrene (PS) to address the need for flexible structures. Nature, on the other hand, has optimized proteins through millions of years of evolution to provide diverse materials with complex structures. Thus studying structural proteins at the molecular level will help us in our endeavor for efficiently designing diverse materials, which are finely tunable, flexible, biodegradable, and have physical properties and functionalities different or superior to those currently present in conventional materials. This will also enable the production of functional devices with less chemical processing and energy consumption, addressing growing environmental concerns.

In this Letter, we studied energy efficiency of whispering-gallery-mode microresonators fabricated solely from semi-crystalline structural proteins. We fabricated microresonators using native SRT (**Fig 1a**), and recombinant SRT (**Fig. 1b**) to demonstrate the possibility of building flexible all-protein functional photonic devices. All-protein microtoroid WGM resonators were fabricated using a molding and solution casting technique (**Fig. S1**). From initial silica templates fabricated as previously described[16], we prepared polydimethylsiloxane (PDMS) negative molds for solution casting of the proteins. The microtoroid structures were then manufactured by filling the respective protein solutions into the corresponding PDMS molds. Light from a tunable laser was evanescently coupled in and out of the fabricated resonators through a tapered optical fiber to measure their resonance properties.

Structural proteins are characterized by long-range ordered molecular secondary structures (e.g., β-sheets, coiled coils, or triple helices) that arise due to highly repetitive primary amino acid sequences within the proteins. These features promote the formation of structural hierarchy via self-assembly. SRT is a semi-crystalline structural proteins, which are flexible, biodegradable, and thermally and structurally stable materials. Semi-crystalline morphology of these proteins, which emanates from their β-sheet secondary structures, is ideal for tunability of their physical properties. SRT protein complex is composed of several proteins varying in molecular weight (e.g., 10-55 kDa) [17]. Although optical properties do not vary with polydispersity in molecular weight due to linear dielectric response or statistically driven size, they may change in special cases due to cooperative non-linear response or thermal effects. Therefore, a recombinant SRT (Rec) protein with unique molecular weight (i.e., 18 kDa) is also selected to create a monodisperse material[18]. A combination of RNA-sequencing, protein mass spectroscopy, and bioinformatics tools (i.e., transcriptome assembly) was performed to produce 18 kDa recombinant SRT protein[18] (**Fig. 1c**). Rec, and SRT proteins were analyzed using X-ray diffraction (XRD), which showed that SRT protein had lower crystallinity compared to Rec protein at room temperature (**Fig. S2a-c**). Rec, and SRT proteins were also analyzed using infrared spectroscopy (FTIR). Amide regions of FTIR data also confirm that the crystalline regions are composed of β-sheets and α-turns (**Fig. S2e-g**). **Figures 1d** present an example protein-based WGM resonator and related transmission spectra of in the 977 nm band. The quality factors of these proteinaceous WGM resonators ranged from $10^5$ to $10^6$ (**Fig. S3**).

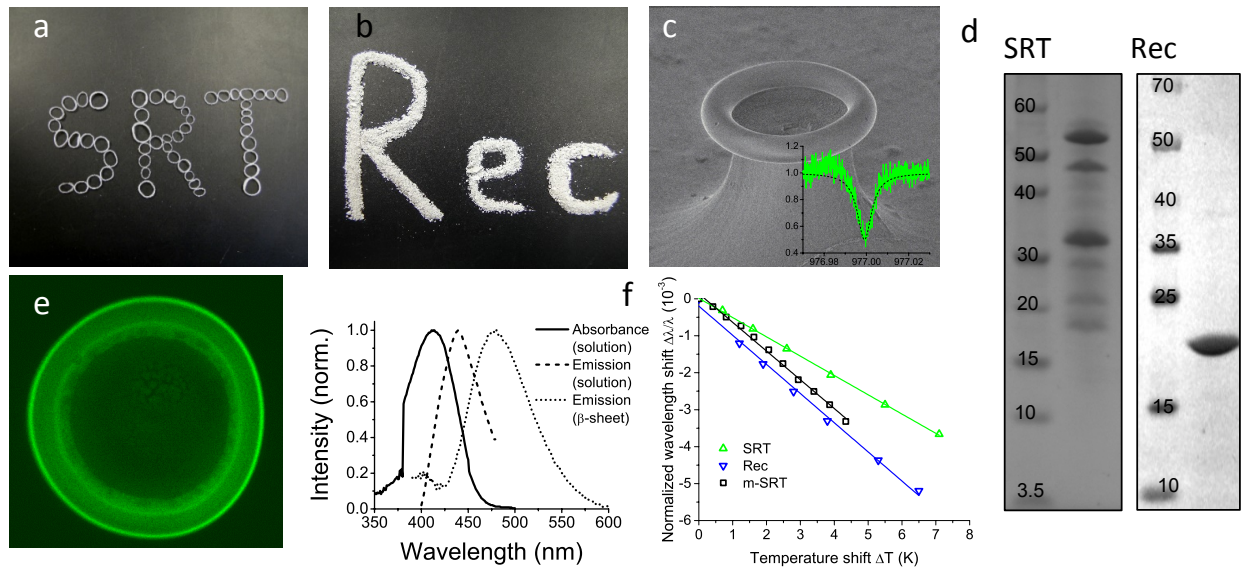

*Figure 1. All-protein whispering-gallery-mode (WGM) microresonators.* (a)-(b) Optical images of SRT rings and recombinant SRT (rec) powder proteins. (c) Scanning electron microscope (SEM) image of a WGM resonator fabricated from SRT protein (Scale bar 20 μm). Insets present typical transmission spectra of WGM resonances in the 977 nm. (d) SDS-Page shows protein molecular weight varying between 15-55 kDA for native SRT, and 18 kDa for recombinant SRT. (e) Green fluorescence emission from an all-SRT chip, which is doped with 1 mM of thioflavin-T (ThT) dye confirming the large red-shift due its binding to β-sheets in the protein. Absorbance and emission spectra of ThT dye. The absorption peak of ThT in solution is at 412 nm. When it is free in a solution, it emits light at 438 nm, and when it is bound to β-sheets in proteins it emits at 482 nm. (f) The temperature dependence of the refractive indices of the respective protein-based WGM resonators made of SRT, recombinant SRT (Rec), and methanol treated SRT (m-SRT).

We probed the nano-crystalline regions in these structural proteins, delineated by anti-parallel β-sheets, using fluorescent microscopy. We selected a fluorescent dye, thioflavin-T (ThT), which binds specifically to the β-sheets, but does not bind to amorphous polypeptides[19]. When ThT dye binds to β-sheets, it undergoes a characteristic red shift of its excitation/emission spectrum (**Fig. 1e**) that may be selectively excited with blue light at 450 nm, resulting in a green fluorescence signal with maximum at 482 and extending to 570 nm. We prepared an all-SRT chip with microresonators, an SRT-coated silica microresonator, a silica resonator partially coated with SRT, and a PDMS chip with microresonators (see Supplementary Information, **Fig. S4**) to test selective binding of ThT dye. The green fluorescence emitted by the all-SRT chip when exposed to blue light is indicative of a large red-shift in emission due to the binding of the ThT dye to β-sheets (**Fig. 1e**).

We experimentally determined the thermo-optic coefficient of SRT proteins by monitoring the resonant wavelength of a WGM resonator while the temperature of the all-protein chip substrate was varied (**Fig. 1f**). We also chemically treated SRT resonator (i.e., sample kept in methanol for 4 hours) and measured the resonant wavelength shift as a function of temperature. For all resonators, we observed that the resonant wavelengths, λ, experience a blue shift (i.e., decrease in wavelength) as the temperature was increased (**Fig. 1f**), i.e., $d\lambda(n,r)/dT < 0$. Subsequently, using the measured parameters in the relation $1/\lambda \; d\lambda(n,r)/dT = 1/n \; dn/dT + 1/r \; dr/dT$, where $T$ is temperature, $n$ is the refractive index, $1/r \; dr/dT$ is the linear thermal expansion coefficient, and $r$ is the radius of the resonator. We measured the thermal expansion coefficient for all protein samples (i.e., $-95 \times 10^{-6} \pm 7 \times 10^{-6}\,K^{-1}$ for SRT proteins), and hence we calculated the normalized thermo-optic coefficients $1/n \; dn/dT$ of SRT and Rec proteins at 1420 nm wavelength.

By coupling proteinaceous WGM resonators to two separate fiber-taper waveguides, we fabricated add-drop filters (**Fig. 2a**) that are frequently integrated in optical communication architectures (such as optical filters, multiplexers and routers), as well as used as schemes for high performance optical sensing devices. As seen in **Fig. 2b**, the add-drop filter routed the light from the input waveguide to the drop port in the second waveguide when the wavelength of the light coincided with the resonance wavelength of the SRT resonator. Non-resonant light passed through the input waveguide to the transmission port. The dip in the transmission port and the peak in the drop port seen in **Fig. 2b** clearly confirm that the SRT add-drop filter performed its function with an add-drop efficiency of 51%, which may be further increased by more careful setting of the waveguide-resonator coupling and by decreasing scattering losses. We also produced proteinaceous waveguides (**Fig. 2c**) using electro-spinning (e-spin) technique (**Fig. S5**). Engaging a single protein fiber to a silica fiber taper via van der Waals attractions, we achieved evanescent coupling of light from the silica fiber taper to the protein fiber, clearly demonstrating the waveguiding capability of proteinaceous fibers. The flexible nature of these protein-based WGM resonators is shown in **Fig. 2d**, where the soft base of the WGM microresonators is bent in a bowed shape. This characteristic enables significant bending on a macroscopic scale (see inset). We performed three-point bending dynamic mechanical analysis (DMA) for the recombinant and native SRT films to quantitatively describe the flexibility of the fabricated photonic microstructures (**Fig. 2d-inset**). The load-deflection curve shows similar bending modulus (~1 GPa) for all films. **Figure 2e** shows bending of the protein substrate as a function of the wavelength of the recombinant SRT resonator. The resonance frequency did not change as the strain in the substrate increased (**Fig. 2e-inset**), which is an important attribute of flexible protein resonators for opto-mechanics applications.

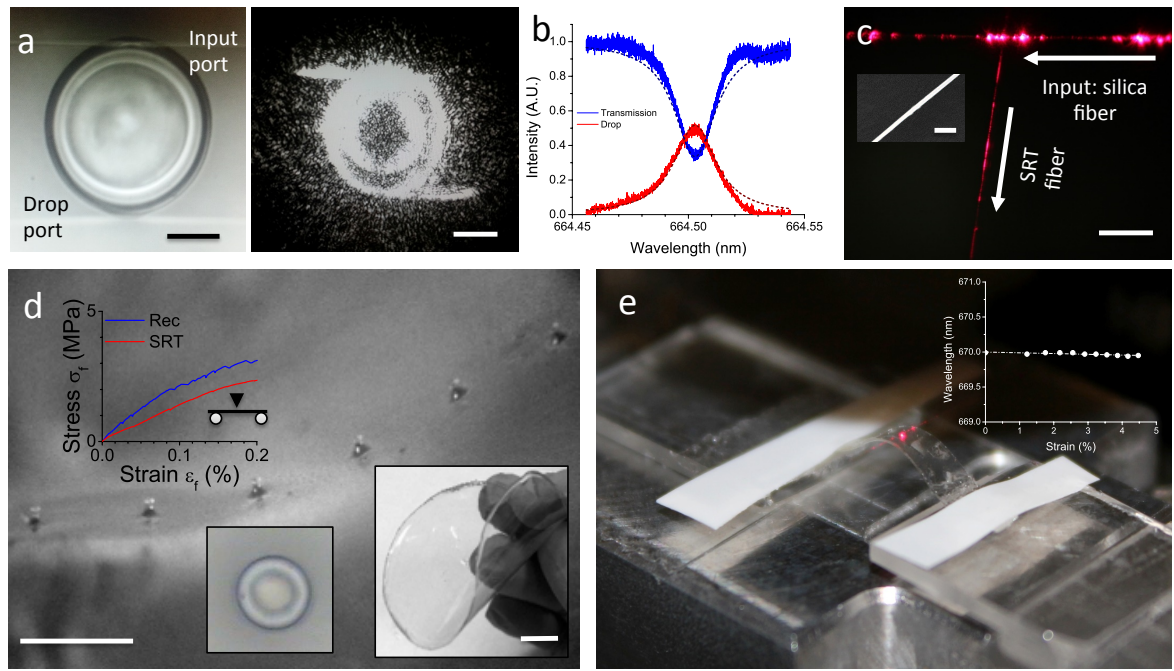

*Figure 2. Device examples of semi-crystalline proteins. (a-b) An all-SRT microtoroid resonator with two coupling tapered-fiber waveguides forming an add-drop filter. Transmission spectrum at 667 nm band shows a resonant dip in the transmission port and a resonance peak in the drop port (scale bars 20 μm). (c) Light transmission through a fiber waveguide prepared from recombinant SRT by electrospinning. Coupling of laser light into the protein fiber was done using a silica tapered fiber (scale bar 50 μm). (d) Flexibility of the substrate (scale bar 1 cm) and of the WGM resonators fabricated from SRT (scale bar 1 mm). Inset presents the results of dynamic mechanical analysis (DMA) measurements used to quantify the bending of native and recombinant SRT films. (e) Wavelength measurements of flexible substrate as a function of strain (inset) show stable resonance in the protein resonator.*

Finally, in order to utilize the strong thermo-optic response of structural proteins, we constructed an all-optical on/off switch using protein-based WGM structure (**Fig. 3a**) where a pump laser (1450nm band) controlled the transmission of a probe light (980 nm band) via the thermal response of the protein. First, two resonance modes were identified (i.e., 1451.7 nm and 974.8 nm). We fixed the wavelength of the probe laser to the probe resonance wavelength where the normalized transmission of the probe along the fiber was set to zero. Then we fixed the wavelength of the pump laser slightly detuned from the pump resonance where the normalized transmission of the pump through the fiber taper was set to unity. Using a square signal, we then tuned the pump laser wavelength to the pump mode wavelength where the normalized transmission of the pump became zero, which in turn heated the cavity and shifted the probe resonance wavelength, effectively moving the probe laser out of resonance with the protein microresonator and the normalized probe transmission became unity, which completed the all-optical switching with an SRT protein microtoroid (**Fig. 3b**). Repeating this experiment for various input powers, we found that the minimum power to perform switching with 12 dB isolation is only 1.6 µW where the circulating power was 143 µW. With silica microtoroid resonators, we obtained 8 dB isolation at 22.75 µW of input power and 223 mW of circulating power. Comparing circulating powers, the protein resonator performs the switching with three orders of magnitude less pump power (**Fig. 3c**). This striking performance is due to the strong thermo-optic coefficients of recombinant SRT protein. Our results show that protein-based photonic devices can be used for low power consumption applications. Moreover, we showed that functional and efficient protein-based photonic devices could be fabricated by tuning their crystallinity at the molecular level, which in turn tunes their thermo-optic response.

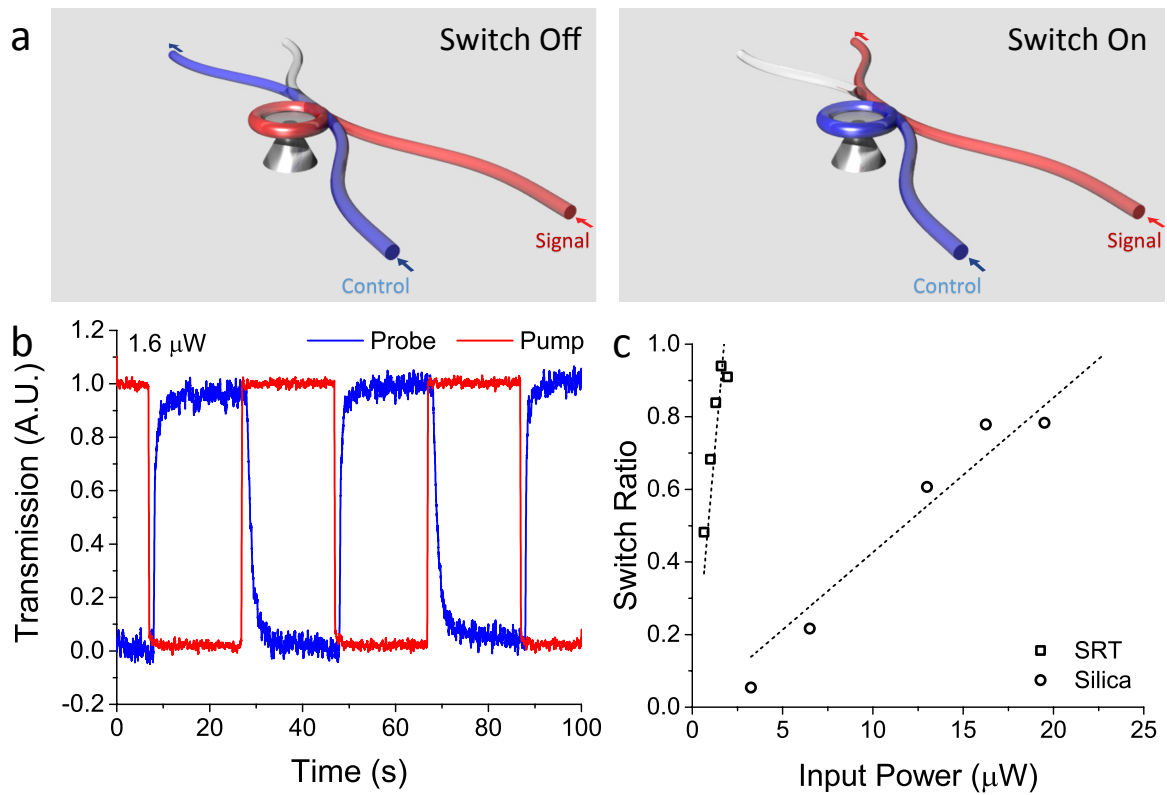

*Figure 3.* All-optical switch based on thermal shift of WGMs in an all-protein microtoroid resonator. (a) Signal in 980 nm band is initially coupled to the resonator where no light transmits and control light in 1450 nm band is just off resonance (left). When the control light is detuned to the microtoroid resonance, cavity temperature increases and the WGM in the 980 nm band becomes off-resonance (right). (b) Normalized transmission of the signal and the control light at 1.6 µW. (c) Ratio of normalized transmission for various input powers. For a silica microtoroid 22.75 µW of control light is required to perform switching.

In summary, proteinaceous WGM microcavities demonstrated here provide a platform to manipulate photons within microscale volumes at high power efficiency. As illustrated here, protein-based materials could provide the next generation of adaptive photonic structures whose optical, mechanical and thermal properties can be engineered at the molecular level for energy efficient applications in photonics. This approach will also expedite the design, fabrication and synthesis of eco-friendly, recyclable, flexible materials and devices.

**Methods**

*Native squid ring teeth (SRT):* European common squid (*Loligo vulgaris*) were caught from the coast of Tarragona (Spain). The squid ring teeth (SRT) were removed from the tentacles, immediately soaked in deionized water and ethanol mixture (70:30 ratio v/v) and vacuum dried in a desiccator.

*Expression and characterization of recombinant squid ring teeth (SRT) proteins:* The SRT protein family is comprised of SRTs of different size that exhibit different physical properties. Heterologous expression of the smallest (~18 kDa) SRT protein extracted from *Loligo vulgaris* (LvSRT) was performed using the protocols described earlier[18]. Briefly, the full length sequence was cloned into Novagen's pET14b vector system and transformed into *E. coli* strain BL21(DE3). Recombinant SRT expression was achieved with a purity of ~90% and an estimated yield of ~50 mg/L. The yield increases approximately ten fold (i.e., ~0.5 g/L) in auto induction media. The size of the protein was confirmed via an SDS-page gel **(Fig. 2d).**

*WGM resonator fabrication:* Details on the Si master preparation and etching for micro-WGM fabrication were reported earlier[20]. Briefly, the microtoroids were fabricated from a 2 µm thick oxide layer on a silicon wafer. First, series of circular pads of 80 µm in diameter were created through a combination of standard photolithography technique and buffered HF etching. These circular pads serve as etch mask for isotropic etching of silicon in $XeF_2$ gas chamber, leaving under-cut silica disks supported by silicon pillar. The microdisks were then selectively reflowed using a 30W carbon dioxide ($CO_2$) laser to form a toroidal shape. A negative mold was made from polydimethylsiloxane (PDMS, Dow Corning Sylgard® 184 silicone elastomer kit). PDMS and curing agent were mixed in a 10:1 ratio, stirred and degased under vacuum for 30 minutes. The mix was poured on the silicon master and it was cured at room temperature for 24 hours. Due to mechanical processing requirements (e.g., viscosity in solvent depends on protein molecular size and solubility), the shape and size of WGM resonators varied. SRT microresonators had a major diameter of 54 μm and a minor diameter of 9 μm, which were used in quality factors and dn/dT measurements. For ThT dye experiments, we used larger SRT microresonators (i.e., 100 μm).

*Protein casting:* SRT solution (either recombinant or native protein) was prepared by dissolving 50 mg/mL of protein in hexafluoro-2-propanol (HFIP). The solution was sonicated for 1 hour and vacuum-filtered in a 4-8 μm mesh size filter. 80 μL of SRT solution were poured into a PDMS toroid mold in successive 20 μL additions 1 minute apart. After the last addition, the HFIP was evaporated at room temperature in the fume hood for 5 minutes and the mold-SRT system was immersed in butadiene (plasticizer) at 80°C (above $T_g$) for 30 minutes. The thermoplastic SRT film was peeled off with tweezers and excess butanediol was removed by rinsing with ethanol.

*WGM / resonance:* Taper-fibers fabricated by heating and pulling single mode fibers were used to couple light from a tunable laser into and out of the microtoroid resonators. The coupling distance between the taper and the resonators was controlled by a nanopositioning system. Four tunable lasers with emission in the wavelength bands of 670 nm, 780 nm, 980 nm and 1450 nm were used to characterize the resonators in different bands of the spectrum. Their wavelengths were linearly scanned around resonances of the resonators. The real-time transmission spectra were obtained by a photodetector connected to an oscilloscope. The scanning speeds and powers of the lasers were optimized in order to eliminate thermally-induced (due to heat build-up in high-Q resonators) line-width broadening and narrowing effects. Typical operating conditions for scanning speed and laser power were 10-20 nm/s and 15 μW, respectively.

*Mechanical characterization:* Protein films were prepared by casting 50 mg/mL SRT/HFIP solution on PDMS molds, resulting in 40x8x0.03 mm rectangular films (length, width, thickness). The samples were analyzed in a dynamic mechanical analysis instrument (TA 800Q DMA) with a three-point bending clamp of 20 mm length between supports. Strain ramp experiments were performed at room temperature at a rate of 250 μm/min with a preload of 0.02 N. The flexural modulus was calculated as described in ASTM D790 – 10.

*Fluorescence imaging*: Thioflavin-T (Sigma) aqueous solution is prepared at 0.125 μM concentration. The absorbance and emission spectra are measured using a fluorescence microscope (Zeiss LSM 5 PASCAL system coupled to a Zeiss Axiovert 200M microscope) at 458 nm wavelength.

*SRT fibers:* Fibers were obtained through an electrospinning process that utilized a conventional horizontal setup from Dr. Seong Kim's laboratory at the Pennsylvania State University. The schematic design consisted of three components (**Fig. S6**): a syringe for supplying the SRT solution, a collecting platform for accumulating the array of fibers, and a voltage source for generating the necessary electric field. A standard procedure began with the preparation of an approximately 50 mg/mL solution of SRT in hexafluoro-2-propanol (HFIP). Once fully dissolved, 0.5 mL of the solution was drawn into the 1 cm diameter plastic syringe, whose 22 gauge bevel tip needle was previously cut perpendicularly to its cylindrical axis, straightening the eventual orientation of the fiber jet. Placing the syringe securely in the pumping device (PHD 2000 Programmable, Harvard Apparatus, Holliston, MA) contained in the ventilation hood, the collecting platform was then configured. Supported by a ring stand and clamps, the collecting platform, constructed from cut pieces of aluminum held 0.8 cm apart by glass slides and epoxy, was positioned such that its center was aligned with the syringe needle, 6.5 cm away from the tip. With the setup complete, the syringe needle was attached to the 10 kV source (HV Power Supply, Gamma High Voltage Research, Ormond Beach, FL), and the aluminum plates of the collecting platform were grounded. After establishing the voltage difference, the pumping device was activated with an infusion rate of 250 μL/min. Following the discharge of approximately 0.25 mL of the solution, the pumping process was terminated and the voltage source detached. To ensure that the HFIP evaporated completely, five minutes were allowed for drying, and the collecting platform was then removed and inspected for fibers between its plates.

**Acknowledgments**

MCD, AP, HJ and RS were supported by the Office of Naval Research under grant No. N000141310595, Army Research Office under grant No. W911NF-16-1-0019, and Materials Research Institute of the Pennsylvania State University. LY, SKO, and HY were supported by the Army Research Office under grant No. W911NF-12-1-0026 and the National Science Foundation under grant No 1264997. We thank Dr. Seong Kim for providing electro-spinning set up.

**Author contributions**

MCD and ŞKÖ conceived the idea and planned the research. MCD derived structure-property equations. MCD, ŞKÖ, and LY supervised the research. HY fabricated the resonators and performed the photonics measurements and data analysis. AP performed the mechanical and spectroscopic measurements and data analysis. RS prepared the fibers and HJ worked on the cloning, recombinant expression and purification of structural proteins. All authors contributed to writing and revising the manuscript, and agreed on the final content of the manuscript.

**Additional information**

The authors declare no competing financial interests. Supplementary information accompanies this paper. Correspondence and requests for materials should be addressed to MCD, ŞKÖ and LY.